\documentclass[aps,pra,twocolumn,superscriptaddress,amsfont,graphicx,,nofootinbib,preprintnumbers]{revtex4-2}
\usepackage{CJKutf8}
\usepackage{color,graphicx,epsfig}
\usepackage{ifpdf}
\usepackage{amsmath}
\usepackage{bm}
\usepackage{color}
\usepackage{xcolor}
\usepackage[english]{babel}
\usepackage{graphicx}
\usepackage{amsfonts}
\usepackage{amssymb}
\usepackage{braket}
\usepackage{hyperref}
\usepackage{xspace}
\usepackage{enumerate}

\usepackage{tikz}
\usetikzlibrary{shapes.geometric,calc}
\usetikzlibrary{decorations.pathmorphing}

\tikzset{
    myvdots/.pic={
        \foreach \i in {1,...,3} {
                \fill[black] (0, -\i * 0.06) circle (0.14ex);
            }
    }
}

\definecolor{nicered}{rgb}{0.7,0.1,0.1}
\definecolor{nicegreen}{rgb}{0.1,0.5,0.1}
\definecolor{kblue}{RGB}{0, 47, 167}
\definecolor{ultramarine}{RGB}{16,9,143}
\definecolor{winered}{RGB}{127,23,52}
\definecolor{nicecyn}{RGB}{0,99,107}
\hypersetup{colorlinks,citecolor= kblue,linkcolor= winered,urlcolor=nicecyn}

\def\beq{\begin{equation}}
\def\eeq{\end{equation}}
\def\bea{\begin{eqnarray}}
\def\eea{\end{eqnarray}}
\def\bsp#1\esp{\begin{split}#1\end{split}}

\newcommand{\rd}{\mathrm{d}}

\newcommand{\coloneq}{:=}

\newcommand{\cN}{\mathcal{N}}

\newcommand{\cA}{\mathcal{A}}

\DeclareMathOperator{\Tr}{Tr}
\DeclareMathOperator{\Sp}{{\textrm{\bf Sp}}}
\DeclareMathOperator{\splitti}{{\textrm{Sp}}}

\begin{document}

\preprint{BONN-TH-2025-21}

\title{Double spacelike collinear limits from multi-Regge kinematics}

\begin{CJK*}{UTF8}{gbsn}

\author{Claude~Duhr}
\author{Aniruddha~Venkata}
\author{Chi~Zhang (张驰)}
\affiliation{Bethe Center for Theoretical Physics, Universität Bonn, D-53115, Germany}

\date{\today}

\begin{abstract}
We study scattering amplitudes and form factors in planar $\mathcal{N}=4$ Super Yang-Mills theory in the limit where two pairs of gluons become collinear. We find that, when the virtualities of both collinear pairs are spacelike, the collinear factorisation of the amplitude involves a generalised splitting amplitude that correlates the two collinear directions, confirming a recent proposal in the literature. Remarkably, we find that our generalised splitting amplitude agrees with the Bern-Dixon-Smirnov (BDS) subtracted six-point amplitude in multi-Regge kinematics. The latter can be explicitly evaluated using integrability to all orders in the coupling. We also present compelling evidence for the universality of the generalised splitting amplitude, by showing that the same function governs the double spacelike collinear limit of scattering amplitudes with up to 8 particles and 3 loops and form factors with up to 6 particles and 2 loops. 
\end{abstract}

\maketitle

\end{CJK*}

\section{Introduction}

It is well known that scattering amplitudes develop singularities whenever subsets of massless partons become collinear. 
From a phenomenological perspective, these collinear singularities govern the parton evolution in the initial and final states. In particular, the evolution of parton distribution functions (PDFs) in Deep Inelastic Scattering (DIS) is described by the celebrated Dokshitzer-Gribov-Lipatov-Altarelli-Parisi (DGLAP) equations~\cite{Gribov:1972ri,Dokshitzer:1977sg,Altarelli:1977zs}, which resum large logarithms resulting from initial-state collinear singularities.
The same PDFs appear in perturbative calculations of differential cross sections for the hadronic process producing a colorless final state $X$~\cite{Collins:1988ig,Collins:1989gx,Collins:2011zzd},
\begin{align}
&\rd \sigma_{h_1h_2\to X} = \label{eq:QCD_fac} \\ 
&\sum_{i,j}\int\rd x_1\,\rd x_2\,f_{i/h_1}(x_1)\,f_{j/h_2}(x_2)\rd\hat{\sigma}_{ij\to X}
 + \mathcal{O}\left(\tfrac{\Lambda_{\textrm{QCD}}}{Q}\right)\,, \nonumber 
\end{align}
where $Q$ is the hard scale of the process and the sum runs over all parton species. The PDF $f_{i/h}(x)$ is the probability density for finding a parton $i$ inside the hadron $h$ carrying a momentum fraction $x$. 
The quantity $\rd\hat{\sigma}_{ij\to X}$ is the partonic differential cross section to produce the final state $X$ from a collision of the parton species $i$ and $j$. The latter quantity exhibits infrared divergences stemming from initial-state collinear radiation, which are absorbed into the renormalisation of the PDFs. The factorisation formula in~\eqref{eq:QCD_fac} is the basis for almost all predictions of observables at hadron colliders, like the Large Hadron Collider (LHC) at CERN. We stress that~\eqref{eq:QCD_fac} has been strictly proven only for final states where all cuts are summed over and sufficient phase space integration has been carried out \cite{Collins:1988ig,Libby:1978qf}.
It is therefore important to understand what are possible limitations of~\eqref{eq:QCD_fac}. Since PDFs are closely tied to initial-state collinear singularities, it is essential to understand how scattering amplitudes factorise in the collinear limit, especially at higher orders in perturbation theory.

In this letter, we study the collinear limit of a scattering amplitude $\cA_N$ for $N$ massless coloured partons. It is well-known that if a subset of partons with momenta $p_i$ and helicities $h_i$, $1\le i\le m<N$, become collinear to a lightlike direction $P$, then the amplitude factorises as
\beq
\cA_N \sim \sum_{c,h_P}\Sp_{h_{P}}^{(m)}\cA_{N-m+1}\,,
\eeq
where `$\sim$' indicates that we neglect terms that are power-suppressed in the limit. The right-hand side involves the hard scattering amplitude $\cA_{N-m+1}$, where the $m$ collinear partons have been replaced by a single parton with momentum $P$, helicity $h_{P}$ and colour $c$.\footnote{The flavour of the parton is fixed uniquely from the flavours of the $m$ collinear partons.} $\Sp_{h_{P}}^{(m)}$ is an operator, called a \emph{splitting amplitude}, that acts on the colour space of the hard amplitude. It only depends on the momenta $p_1, \ldots,p_m$ from the collinear set. Moreover, if the splitting is timelike (meaning that $(p_1+\ldots+p_m)^2>0$), then also the colour degrees of freedom factorise, and $\Sp_{h_{P}}^{(m)}$ only depends on the colour charges of the collinear particles  (cf.~\cite{Kosower:1999xi,Feige:2014wja} for a proof of the factorisation). For spacelike splittings, however, the colour degrees of freedom may not completely factorise, and $\Sp_{h_{P}}^{(m)}$ may depend on the colour charges of the particles \cite{Catani:2011st,Henn:2024qjq,Guan:2024hlf}. Splitting amplitudes are known up to $m=4$ at  tree-level, $m=3$ and one-loop and for $m=2$ up to three loops~\cite{Catani:2003vu,Duhr:2014nda,Badger:2015cxa,Czakon:2022fqi,Henn:2024qjq,Guan:2024hlf}.

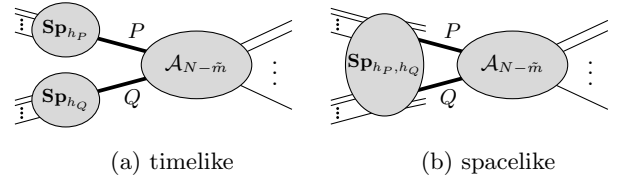
\begin{figure}
    \begin{center}
        \begin{tikzpicture}
            \begin{scope}
                 \coordinate (O) at (0,0);
                \path (O) +(180-15:1.8) coordinate (A);
                 \path (O) +(15-180:1.8) coordinate (B);
                \draw[line width=0.5mm] (O)--(A);
                \draw[line width=0.5mm] (O)--(B);
                  \draw (O) -- (25:1.4);
                  \draw ([yshift=-1ex]O) -- +(25:1.4);
                    \draw (O) -- (-25:1.4);
                \node at([xshift=3.2em]O) {$\vdots$};
                \draw ([yshift=1pt]A)--+(180-15:0.7);
                \draw ([yshift=3pt]A)--+(180-15:0.7);
                \draw ([yshift=-6pt]A)--+(180-15:0.7);
                 \draw ([yshift=3pt]B)--+(15-180:0.7);
                \draw ([yshift=5pt]B)--+(15-180:0.7);
                \draw ([yshift=-4pt]B)--+(15-180:0.7);
                \pic at ([xshift=-1.8em,yshift=1.4ex]A) {myvdots};
                 \pic at ([xshift=-1.8em,yshift=-0.4ex]B) {myvdots};
                 \node[scale=0.9] at (-0.8,0.45) {$P$};
                 \node[scale=0.9] at (-0.85,-0.45) {$Q$};
                \node[ellipse, fill=gray!30,minimum height=1cm, draw,scale=0.9] at(O) {$\cA_{N-\tilde{m}}$};
                \node[ellipse,fill=gray!30,draw,scale=0.9,minimum height=0.8cm,minimum width=1.0cm] at (A) {};
                \node[scale=0.8] at(A) {$\Sp_{h_{P}}$};
                \node[ellipse,fill=gray!30,draw,scale=0.9,minimum height=0.8cm,minimum width=1.0cm] at (B) {};
                \node[scale=0.8] at(B) {$\Sp_{h_{Q}}$};
                \node at([xshift=-1em,yshift=-4em]O) {(a) timelike};      
            \end{scope}
             \begin{scope}[shift={(4.2,0)}]
                 \coordinate (O) at (0,0);
                \path (O) +(180-15:1.8) coordinate (A);
                 \path (O) +(15-180:1.8) coordinate (B);
                \draw[line width=0.5mm] (O)--(A);
                  \draw[line width=0.5mm] (O)--(B);
                  \draw (O) -- (25:1.4);
                  \draw ([yshift=-1ex]O) -- +(25:1.4);
                    \draw (O) -- (-25:1.4);
                \node at([xshift=3.2em]O) {$\vdots$};
                      \node[scale=0.9] at (-0.8,0.45) {$P$};
                 \node[scale=0.9] at (-0.85,-0.45) {$Q$};
                \draw ([yshift=5pt]A) -- +(-12:0.6);
                \draw ([yshift=3pt]A) -- +(-12:0.6);
                \draw ([yshift=-5pt]B) -- +(12:0.6);
                \draw ([yshift=-3pt]B) -- +(12:0.6);
              \draw ([yshift=1pt]A)--+(180-15:0.7);
                \draw ([yshift=3pt]A)--+(180-15:0.7);
                \draw ([yshift=-6pt]A)--+(180-15:0.7);
                 \draw ([yshift=3pt]B)--+(15-180:0.7);
                \draw ([yshift=5pt]B)--+(15-180:0.7);
                \draw ([yshift=-4pt]B)--+(15-180:0.7);
                \pic at ([xshift=-1.8em,yshift=1.4ex]A) {myvdots};
                 \pic at ([xshift=-1.8em,yshift=-0.4ex]B) {myvdots};
                \node[ellipse, fill=gray!30,minimum height=1cm, draw,scale=0.9] at(O) {$\cA_{N-\tilde{m}}$};
                \node[ellipse,fill=gray!30,draw,scale=0.8,minimum height=1.7cm,minimum width=1.3cm] at (-1.7,0) {};
                \node[scale=0.8] at(-1.7,0) {$\Sp_{h_{P},h_{Q}}$};
                 \node at([xshift=-1em,yshift=-4em]O) {(b) spacelike};
            \end{scope}
        \end{tikzpicture}
    \end{center}
    \caption{Timelike versus spacelike collinear factorisation where $\tilde{m}=N-m-m'+2$. The two heavy lines represent the parent partons with momenta and helicities $(P, h_P)$ and $(Q, h_Q)$, respectively. Nearly parallel lines on the same side of the bubble correspond to the timelike collinear limit, while those on opposite sides correspond to the spacelike collinear limit}
   \label{fig:SL_and_TL_split}
\end{figure}

So far the discussion was restricted to the case of a subset of partons becoming collinear to the same direction $P$. In the case of multiple timelike collinear directions, the factorisation involves a product of splitting amplitudes (cf.~fig.~\ref{fig:SL_and_TL_split}(a)), one for each collinear direction. 
It was recently proposed that, in the case of two spacelike collinear directions $P$ and $Q$, collinear factorisation involves a \emph{generalised splitting amplitude} that depends on the combined set of quantum numbers from both collinear sets~\cite{Cieri:2024ytf} (cf.~fig.~\ref{fig:SL_and_TL_split}(b)),
\beq\label{eq:DSL_coll_fac}
\cA_N \sim \sum_{c,h_{P},c',h_{Q}}\Sp_{h_{P},h_{Q}}^{(m_1,m_2)}\cA_{N-m_1-m_2+2}\,,
\eeq
where $\Sp_{h_{P},h_{Q}}^{(m_1,m_2)}$ is the generalised splitting amplitude from~\cite{Cieri:2024ytf}.
Equation~\eqref{eq:DSL_coll_fac} implies that the two spacelike collinear clusters do not completely factorise from each other, but there are correlations involving both colour and kinematics between the two collinear sets. In particular,~\eqref{eq:DSL_coll_fac} may lead to a breakdown of~\eqref{eq:QCD_fac} for some observables, because~\eqref{eq:QCD_fac} does not account for collinear singularities that correlate the two incoming beams. 

Up to now there has been no study of generalised splitting amplitudes. It is easy to see that correlations between two spacelike collinear sets require loop amplitudes with many legs.  In a general gauge theory, we do currently not have access to results for amplitudes with more than 5 legs and 2 loops. The situation is very different in the planar limit of the maximally supersymmetric $\mathcal{N}=4$ Super Yang-Mills (SYM) theory, where analytic results for amplitudes with many loops and legs are available (cf.,~e.g.,~\cite{Caron-Huot:2019vjl,Dixon:2016nkn,Drummond:2018caf,Caron-Huot:2011zgw,He:2019jee,He:2020vob,Li:2021bwg}). In the remainder of this paper we study scattering amplitudes in planar $\cN=4$ SYM in the limit where two spacelike pairs of massless particles become collinear. This will allow us to test for the very first time the proposal of~\cite{Cieri:2024ytf}. Moreover, we will show that in this case there is a duality between the double spacelike (DSL) collinear limit and another kinematic limit, namely the so-called multi-Regge limit of the 6-point amplitude.  Remarkably, the latter can be described to all order from integrability~\cite{Basso:2014pla}, and the duality between the two limits implies that we can compute our generalised splitting amplitude to all orders in the coupling.

\section{Amplitudes in planar $\cN=4$ SYM}
Amplitudes in planar $\cN=4$ SYM are well studied (see~\cite{Arkani-Hamed:2022rwr} for a review). Many analytic results for multi-leg multi-loop amplitudes are available, due to the large number of symmetries of the theory. Besides the usual superconformal symmetry, the $S$-matrix of planar $\cN=4$ SYM is invariant under a dual superconformal symmetry~\cite{Drummond:2006rz}, which closes with the ordinary superconformal symmetry into a Yangian algebra~\cite{Drummond:2009fd}. It is convenient to consider superamplitudes, where each external state is characterised by a lightlike momentum $p_i$, an adjoint colour index $a_i$ and a Graßmann variable $\eta_i$ transforming in the fundamental representation of the $R$-symmetry group SU$(4)$~\cite{Nair:1988bq,Arkani-Hamed:2008owk}. In the planar limit only single-trace colour structures contribute, so that the superamplitude reads
\beq
\cA_N= \!\!\!\sum_{\sigma\in S_N/\mathbb{Z}_N}\!\!\!\Tr(T^{a_{\sigma_1}}\cdots T^{a_{\sigma_N}})A_N(\Lambda_{\sigma_1},\ldots,\Lambda_{\sigma_N})\,,
\eeq
where $T^a$ are the generators of the gauge group SU$(N_c)$, $A_N$ are the partial amplitudes that depend on the cyclic ordering of the particles and $\Lambda_i\coloneq(p_i,\eta_i)$, but are independent of the colour. As a consequence of dual (super-) conformal symmetry, the partial amplitude can be cast in the form~\cite{Drummond:2008vq}
\beq
A_N({\bf \Lambda}_N) = A_{\textrm{MHV}}^{\textrm{tree}}({\bf \Lambda}_N)\,A_N^{\textrm{BDS}}\,R_N({\bf \Lambda}_N)\,,
\eeq
where ${\bf \Lambda}_N \coloneq (\Lambda_1,\ldots,\Lambda_N)$, $A_{\textrm{MHV}}^{\textrm{tree}}$ is the maximally helicity violating (MHV) tree superamplitude, and $A_N^{\textrm{BDS}}$ is the BDS amplitude, defined as the exponential of the one-loop MHV amplitude (divided by the tree-level) weighted by the cusp and collinear anomalous dimensions~\cite{Bern:2005iz}.
The BDS amplitude captures infrared singularities to all orders in perturbation theory, so that the \emph{BDS-subtracted amplitude} $R_N$ is finite and dual conformally invariant~\cite{Drummond:2008vq,Bern:2008ap}. In particular, its kinematic dependence can only be through dual conformally invariant cross-ratios, 
\beq
u_{ijkl} = \frac{x_{ij}^2x_{kl}^2}{x_{ik}^2x_{jl}^2}\,,\quad x_{ij} = x_i-x_j=p_{i}+\cdots+p_{j-1}\,,
\eeq
where the $x_i$ are the dual coordinates; therefore, $R_N$ is non-trivial only for $N\ge 6$. Explicit analytic results for BDS-subtracted amplitudes are available both for MHV and next-to-MHV (NMHV) for $N=6$ up to $7$ and  $6$ loops~\cite{Caron-Huot:2019vjl}, for $N=7$ up to $4$ loops~\cite{Dixon:2016nkn,Drummond:2018caf}, and for $N=8$ up to 3 and 2 loops~\cite{Li:2021bwg,He:2019jee}, respectively. The symbols of all master integrals for all two-loop amplitudes in $\mathcal{N}=4$ SYM have been computed in~\cite{Caron-Huot:2011zgw,Spiering:2024sea}.

Let us briefly comment on the behavior of $A_N$ in the limit where a pair of particles $i$ and $j$ become collinear. First of all, it is easy to see that the leading power contribution only occurs when two collinear particles are adjacent in the colour ordering. 
Moreover, since $R_N$ is unity or an $R$-invariant~\cite{Mason:2009qx} for $N\le 5$, it follows that the 2-particle splitting amplitude is entirely determined by the BDS ansatz~\cite{Anastasiou:2003kj,Bern:2005iz}. For example, in the collinear limit $p_1 \parallel p_2$, an MHV component amplitude factorises as
\beq\label{eq:split_2}
A_{5}^{\text{MHV}}(p_{1}^{h_{1}},p_{2}^{h_{2}},\ldots) \sim \splitti_{+}^{(2)}(p_{1}^{h_{1}},p_{2}^{h_{2}}; \xi)\,A_{4}(P^{h_{P}},\ldots)\,,
\eeq
where the subscript on $\splitti$ now denotes its MHV degree (which can be $\pm$, requiring $h_{1}+h_{2}-h_{P}=\pm1$), and $\xi$ is the fraction of the parent momentum $P$ carried by $p_1$, i.e., $p_1=\xi P$ in the strict collinear limit. Since the splitting amplitude is completely determined by $N=5$, the BDS-subtracted amplitude $R_6$ must tend to its tree value
in all two-particle collinear limits. 
Using the fact that amplitudes in this theory are  
dual to Wilson loops~\cite{Caron-Huot:2010ryg}, it is possible to compute a systematic expansion of the BDS-subtracted amplitudes around collinear limits. The quantities that control this near-collinear expansion can be computed to all orders using tools from integrability~\cite{Basso:2013vsa}.

\section{The generalised splitting amplitude} \label{sec: general splitting amplitude}
We now discuss the main result of this letter, namely the generalised splitting amplitude for two adjacent spacelike collinear pairs, say $p_{1}\parallel p_{2}$ and $p_{3}\parallel p_{4}$, in planar $\cN=4$ SYM to all orders in the 't Hooft coupling constant $a$. Note that the generalised splitting amplitudes, which require $P\cdot Q\coloneq(p_{1}+p_2)\cdot (p_{3}+p_{4}) \neq 0$, 
appear starting from $N=6$.

Before presenting the result, we first clarify what is meant by `spacelike'. 
Amplitudes in planar $\cN=4$ SYM are typically computed in the Euclidean region where all invariants $x_{ij}^2$  are negative and the amplitude is real. We refer to the region where all invariants are timelike, i.e., positive, as the \emph{timelike Euclidean region}. It is identical to the Euclidean region up an overall phase (and the BDS-subtracted amplitude is independent of this phase). In the following, the term `spacelike' refers to analytic continuation in some two-particle invariants $x^2_{i,i+2}$
from the timelike Euclidean region  to the region where these invariants are negative. 

We are now in a position to state our claim about the generalised splitting amplitude, which consists of two points: (i) The 
splitting amplitudes factorise in \emph{all but two} distinct double spacelike collinear regions,  
which, in addition to satisfying $s_{12}=x_{13}^2 < 0$ and $s_{34}=x_{35}^2 < 0$, require 
\begin{align}
    \text{(I)}:&\quad P\cdot Q>0,\,s_{23}=x^2_{24}>0,\,\xi_{1}>1,\,\xi_{2}>1\:, \label{eq:conlinear_region_I}\\
    \text{(II)}:&\quad P\cdot Q<0,\,s_{23}=x^2_{24}<0,\,\xi_{1}>1,\,\xi_{2}>1\:,\label{eq:conlinear_region_II}
\end{align}
respectively, and $\xi_{1}$ and $\xi_{2}$ are the momentum fractions of $p_{1}$ and $p_{4}$ in the collinear limit. 
(ii) Although the (component) amplitudes in these two regions do not factorise, they are compensated by \emph{universal} corrections,
\begin{align}\label{eq:split_A6}
\nonumber A_N &\sim \sum_{h,h'}\splitti^{(2)}_{h}(-p_1,p_2;\xi_1+i\varepsilon)\,\splitti_{h'}^{(2)}(\pm p_3,\mp p_4;\xi_2+i\varepsilon)\,\\
&\quad\,\times \Delta^{h,h'}_{\rm{I/II}}(\tau,z) A_{N-2}(P^{h_{P}},Q^{h_{Q}},\ldots)\,,
\end{align}
where two choices of signs correspond to two regions, with the $+$ or $-$ sign of the momentum indicating that it is outgoing or incoming, respectively. Here, $\splitti^{(2)}_{h}$ is the 2-particle splitting amplitude defined in eq.~\eqref{eq:split_2}, which is determined by the BDS amplitude. The new ingredient $\Delta_{\rm{I/II}}^{h,h'}$ receives contributions from the BDS-subtracted amplitudes and is completely determined by the multi-Regge limit of the six-point BDS-subtracted amplitude $R_{6}$, as we will demonstrate below.

\subsection{The six-point case}

To analyze the amplitudes in the collinear regions \eqref{eq:conlinear_region_I} and \eqref{eq:conlinear_region_II}, we need to first analytically continue the amplitudes to the kinematic regions defined in~\cite{Bartels:2008ce,Bartels:2009vkz} (where they are called \emph{Mandelstam regions}) containing these collinear regions, and then take $x_{13}^2$ and $x_{35}^2$ to 0 at the same rate.\footnote{\label{fn_Supp_Material}The details of how the limit is parametrised can be found in the App.~\ref{App:a}.} For simplicity, we present only the analysis for region \eqref{eq:conlinear_region_I} here, as other regions follow similarly.

For general $N$, there may be  multiple regions containing the collinear region of interest. 
However, there is only one physical region containing region \eqref{eq:conlinear_region_I} for $N=6$, corresponding to the scattering $p_{1}p_{4}\to p_{2}p_{3}p_{5}p_{6}$:
\begin{align}\label{eq:collinear_region_for_hexagonI}
    x_{24}^2 > 0, \quad x_{15}^2 > 0, \quad \text{and all other } x_{ij}^2 < 0\,.
\end{align}
There are 3 independent conformally-invariant cross-ratios for $N=6$: 
\beq
u_1 = u_{1346}\,,\quad u_2=u_{2451}\,,\quad u_3 = u_{3562}\,.
\eeq
The DSL collinear limit then corresponds to analytically continuing $u_2\to u_2 e^{2\pi i}$ and then taking the limit $(u_1,u_2,u_3)\to (0,1,0)$, with the following ratios held fixed:
\beq
\frac{u_1}{1-u_2}=\frac{z\bar{z}}{(1-z)(1-\bar{z})} \,,\quad \frac{u_3}{1-u_2}= \frac{1}{(1-z)(1-\bar{z})} \,.
\eeq
Remarkably, at the level of the cross-ratios, the DSL collinear limit is \emph{identical} to the multi-Regge limit for the six-point BDS-subtracted amplitude, where the produced particles are strongly ordered in rapidity. It was first observed in~\cite{Bartels:2008ce} that the BDS-subtracted amplitude $R_6$ in the Euclidean region becomes trivial in multi-Regge kinematics (MRK), but the result is non-trivial if continued to the Riemann sheet defined by~\eqref{eq:collinear_region_for_hexagonI} due to the presence of a Regge cut. Since both the analytic continuation and the limit agree for the DSL collinear limit and MRK at the level of cross-ratios, we can immediately identify the generalised splitting amplitude $\Delta_{\rm I}^{h,h'}(\tau\coloneq\sqrt{u_{1}u_{3}},z)$ in eq.~\eqref{eq:split_A6} with various six-point BDS-subtracted amplitudes in MRK. 

More precisely, we find that $\Delta_{\rm I}^{++}=\Delta_{\rm I}^{--}$ can be obtained from the MHV BDS-subtracted amplitude $R_{6}^{\rm{MHV}}$ in MRK~\cite{Bartels:2008ce,Bartels:2009vkz,Lipatov:2010ad,Fadin:2011we,Caron-Huot:2013fea}, and $\Delta_{\rm I}^{+-}=\Delta_{\rm I}^{-+}\vert_{z\leftrightarrow\bar{z}}$ from the NMHV BDS-subtracted amplitude $R_{6}^{\rm{NMHV}}$ in MRK~\cite{Lipatov:2012gk,Dixon:2014iba}~(For more details, see App.~\ref{App:b}). For example, from an MHV configuration in $2\to 4$ kinematics, we have (cf. \cite{DelDuca:2018hrv}),
\begin{align} \label{eq:Delta_RMRK} 
& \Delta_{\rm I}^{++}(\tau,z) = e^{i\delta_6}\mathcal{M}(R_6^{\textrm{MHV}})= \cos\pi\omega_{\textrm{ab}}  \\
&+\tfrac{ia}{2}\!\sum_{n=-\infty}^{+\infty}(-1)^{n}\left(\tfrac{z}{\overline{z}}\right)^{n/2}\mathcal{P}\int_{-\infty}^{+\infty}\frac{|z|^{2i\nu}\Phi^{\textrm{reg}}_{\nu n}\,\rd\nu}{\nu^2+ n^2/4 }\left(-\tau\right)^{-\omega_{\nu n}},  \nonumber
\end{align}
where $\mathcal{P}$ represents the principal value integral, and $\mathcal{M}(R)$ indicates that the function $R$ is evaluated in MRK.
Here $\omega_{\textrm{ab}}=\tfrac{1}{8}\gamma_K(a)\log|z|^2$ is the Regge pole contribution, and $\gamma_K(a)$ is the cusp anomalous dimension. The remaining contribution takes the form of a Fourier-Mellin transform and involves the BFKL eigenvalue $\omega_{\nu n}$ and the (regularised) impact factor $\Phi^{\textrm{reg}}_{\nu n}$. Similarly, $\Delta_{\rm II}(\tau,z)$ for the region \eqref{eq:conlinear_region_II} is determined by $e^{-i\delta_{6}}\mathcal{M}(R_{6}^{\textrm{MHV}})$ in $3\to 3$ kinematics~\cite{Lipatov:2010ad,Bartels:2010tx,Basso:2014pla}, which has an expression similar to eq.~\eqref{eq:Delta_RMRK}.

Our analysis reveals a duality between the DSL collinear limit and MRK in planar $\cN=4$ SYM. 
The fact that the BDS-subtracted amplitude vanishes in Euclidean MRK corresponds to the fact that the amplitude factorises into 2-particle splitting amplitudes in the double timelike collinear limit. Correspondingly, the non-factorising contributions in the DSL collinear limit can be attributed to the same Glauber gluon exchange~\cite{Sen:1982xv,Collins:2011zzd} described by the Fourier-Mellin integral in eq.~\eqref{eq:Delta_RMRK}. We note that both kinematic regimes explore aspects of forward scattering. Nevertheless, the two regimes are distinct, as no rapidity ordering is imposed in the collinear limit, making the duality between the collinear limit and MRK non-trivial.

As a consequence of the conformal equivalence of the two limits, we can directly use knowledge about $R_6$ in MRK and apply it to the DSL collinear limit. In particular, we see that $\Delta_{\rm I/II}$ is completely determined by the cusp anomalous dimension $\gamma_K$, the BFKL eigenvalue $\omega_{\nu n}$ and the impact factor $\Phi^{\textrm{reg}}_{\nu n}$, where the latter two quantities have been extracted to high orders from fixed-order computations~\cite{Lipatov:2010ad,Fadin:2011we,Dixon:2012yy,Dixon:2014voa}. More importantly, since all three quantities are known to all orders from integrability~\cite{Beisert:2006ez,Basso:2014pla} (at least conjecturally), we conclude that also $\Delta_{\rm I/II}$ is fixed to all orders in the 't Hooft coupling. It is known how to systematically evaluate the Fourier-Mellin transform in eq.~\eqref{eq:Delta_RMRK} order by order in terms of Brown's single-valued harmonic polylogarithms~\cite{Remiddi:1999ew,BrownSVHPLs,brownSV}, and so we conclude that the same statement must hold for $\Delta_{\rm I/II}$.

\subsection{Universality of generalised splitting amplitudes}

At this point, we would like to check that $\Delta_{\rm I/II}$ is universal. It is independent of the hard scattering and is the same function for all higher point amplitudes and form factors.  We now give compelling evidence that it is.

For the higher-point amplitudes, the main difference to the six-point case is that there may be several physical regions compatible with either DSL collinear region \eqref{eq:conlinear_region_I} or \eqref{eq:conlinear_region_II}. For instance, physical regions compatible with both collinear regions can have either $x_{61}^2>0$ or $x_{61}^2<0$ for $N=7$. Through analytic continuation to various compatible regions, we have performed explicit computations of the DSL collinear limits for the symbols of seven- and eight-point MHV and NMHV amplitudes up to 3 loops~\cite{Caron-Huot:2010ryg,Drummond:2014ffa,Dixon:2016nkn,Li:2021bwg}. Remarkably, we find that all these amplitudes exhibit the same pattern as in \eqref{eq:split_A6}, where $\tau$ and $z$ are defined as in the six-point case with $p_6$ replaced by $p_N$.

The four-point MHV $\Tr \phi^2$ form factor is the first to exhibit non-trivial DSL collinear limits. This form factor has been computed through three-loop order~\cite{Dixon:2022xqh,Dixon:2024yvq}, while the five- and six-point MHV form factors have been computed through two-loop order~\cite{Li:2024rkq}. We have verified that all these form factors exhibit the same DSL collinear limit behaviour (with the obvious identification $x_{13}^2=s_{12}\coloneq (p_1+p_2)^2$, $x_{24}^2=s_{23}$, etc.) as the amplitudes.
In particular, the four-point form factor fails to factorise not only in regions I and II described in \eqref{eq:conlinear_region_I} and \eqref{eq:conlinear_region_II}, but also in regions $\rm{I}'$ and $\rm{II}'$, which are related to regions I and II via $\xi_i \to 1-\xi_i$, due to an additional symmetry. Nevertheless, the form factor exhibits the same pattern as in eq.~\eqref{eq:split_A6} in all aforementioned regions with the correct identification of $\tau$ and $z$.

Based on our explicit computations, we conjecture that the generalised splitting amplitude is universal, and that the function $\Delta_{\rm I/II}$ governs the behavior of any quantity in planar $\cN=4$ SYM in the limit where 2 spacelike pairs of particles become collinear.

\section{Conclusions}
In this paper we have obtained for the first time results for high-loop scattering amplitudes in the collinear limit where two spacelike pairs of massless particles become simultaneously collinear, and we have obtained an analytic expression for a generalised splitting amplitude, first introduced in~\cite{Cieri:2024ytf}.

We have focused our study on the planar $\cN=4$ SYM theory.
We have uncovered a duality between the DSL collinear limit and MRK for the six-point BDS-subtracted amplitude. MRK amplitudes in planar $\cN=4$ SYM have been studied extensively, both for $N=6$~\cite{Bartels:2008ce,Caron-Huot:2013fea,Bartels:2009vkz,Bartels:2010tx,Fadin:2011we,Lipatov:2010ad,Dixon:2012yy,Pennington:2012zj,Lipatov:2012gk,Dixon:2014voa,Basso:2014pla,Broedel:2015nfp,Bartels:2011xy} and beyond~\cite{Bartels:2011ge,Lipatov:2010qg,Prygarin:2011gd,Bartels:2012gq,Bartels:2013jna,Bartels:2014jya,Bartels:2014ppa,Bartels:2014mka,Bargheer:2015djt,Broedel:2016kls,Bargheer:2016eyp,DelDuca:2016lad,DelDuca:2018hrv,Marzucca:2018ydt,DelDuca:2019tur,DelDuca:2018raq,Bargheer:2019lic,Bartels:2020twc}. Our duality allows us to immediately apply known results for the 6-point BDS-subtracted amplitude in MRK to obtain the generalised splitting amplitude $\Delta_{\rm I/II}$. The universality of the DSL collinear limit can be used to turn the MRK results for $R_6$ into constraints for the a future bootstrap program for higher-point amplitudes and forms factors. It would be interesting to understand if there are also kinematic regimes for $N>6$ that are dual to the BDS-subtracted amplitudes $R_N$ in MRK.

Our findings confirm the expectations of~\cite{Cieri:2024ytf} about the structure of the DSL collinear limit. 
Unlike other sources of collinear factorisation breaking considered in the literature~\cite{Catani:2011st,Henn:2024qjq}, this correlation between different spacelike collinear directions contributes already at leading color. While our analysis was restricted to the planar $\cN=4$ SYM theory, this is sufficient to establish the existence of correlations between  multiple spacelike collinear directions at leading color. For the future, it would be tantalising to explore the implications for the factorisation of observables in hadronic collisions into PDFs multiplied by partonic cross sections.

\begin{acknowledgments}
 We are grateful to Zhenjie Li for his instruction regarding the higher-point form factors.
This research was supported by 
the European Research Council (ERC) under the European Union’s research and innovation programme grant agreement 101043686 (ERC Consolidator Grant LoCoMotive).
Views and opinions expressed
are however those of the author(s) only and do not necessarily reflect those of the
European Union or the European Research Council. Neither the European Union
nor the granting authority can be held responsible for them.

\end{acknowledgments}

\appendix

\section{Parametrisation of the double collinear limit} \label{App:a}

In this appendix, we describe how to parametrise the double collinear limit using momentum twistors~\cite{Hodges:2009hk}, which are convenient for the evaluation of amplitudes and form factors in $\mathcal{N}=4$ SYM.

For amplitudes, the momentum twistors are defined as $Z_{i}\coloneq(\lambda_{i}^{\alpha},x^{\alpha\dot{\alpha}}_{i}\lambda_{i\alpha})$ where $\lambda_{i}$ are the usual spinor-helicity variables and $x_{i}^{\alpha\dot{\alpha}}\coloneq x_{i}^{\mu}\sigma_{\mu}^{\alpha\dot{\alpha}}$. The double collinear limit can then be parametrised in terms of momentum twistors as
\begin{align*}
    Z_{2} &\to Z_{1}+\epsilon Z_{4}-\epsilon\tau_{1}\frac{\langle 14\rangle}{\langle N1\rangle}Z_{N}
    -\epsilon^{2}(1-z)\tau_{2}\frac{\langle 14\rangle}{\langle 45\rangle}Z_{5} \,,\\
    Z_{3} &\to Z_{4}+\epsilon Z_{1}-\epsilon\tau_{2}\frac{\langle 14\rangle}{\langle 45\rangle} Z_{5} 
    +\epsilon^2 \frac{(1-\bar{z})\tau_{1}}{\bar{z}}\frac{\langle 14\rangle}{\langle N1\rangle}Z_{N} \,.
\end{align*}
Here, the collinear limit is taken by letting $\epsilon \to 0$, and the parameters $\tau_i$ are related to the momentum fractions $\xi_i$ through $\tau_i = (1 - \xi_i)/\xi_i$. 

Compared with amplitudes, the momentum twistors for $N$-point form factors additionally have $Z_{N+i}=\mathsf{P}Z_{i}$ with the transition matrix
\begin{equation*}
    \mathsf{P}= \begin{pmatrix}
        \delta_{\beta}^{\alpha} & 0 \\
        q_{\beta}{}^{\dot{\alpha}} & \delta_{\beta}{}^{\dot{\alpha}}
    \end{pmatrix} \:,
\end{equation*} 
such that $x_{N+i}-x_{i}=q$ (note that $q=0$ results in the amplitude case). For a comprehensive study, refer to e.g.~\cite{Basso:2023bwv}.

\section{More details on the six-point case} \label{App:b}

In this appendix, we present more details on the DSL collinear limit for the six-point amplitudes.
Without loss of generality, we set the helicities of particles 1 through 4 as $\{-,+,+,-\}$, with particles 5 and 6 arbitrary (where all particles are taken to be outgoing).
In the DSL collinear limit $p_{1}\parallel p_{2}$ and $p_{3}\parallel p_{4}$, 
this amplitude behaves, according to eq.~\eqref{eq:DSL_coll_fac}, as
\begin{align}
    &A(1^{-},2^{+},3^{+},4^{-},5^{h_{5}},6^{h_{6}}) \to  \\
   & \sum_{h,h'}\splitti^{(4)}_{h,h'}(p_{1},p_{2},p_{3},p_{4};\xi_{1},\xi_{2})
    A(P^{h_{P}},5^{h_{5}},6^{h_{6}},Q^{h_{Q}}) \:, \nonumber
\end{align}
where the superscript $h_{i}$ denotes the helicity of particle~$i$, $h=h_{1}+h_{2}-h_{P}$ and $h'=h_{3}+h_{4}-h_{Q}$ are the MHV degree of the two collinear pairs, while $\xi_{1,2}$ denote the momentum fractions in the collinear limit through $p_{1}=\xi_{1}P$ and $p_{4}=\xi_{2}Q$.

We first consider Region I, corresponding to the scattering process $1^{+}4^{+}\to 2^{+}3^{+}5^{h_{5}}6^{h_{6}}$. (Note that particle helicities are flipped when converting from outgoing to incoming, but the MHV degree of amplitudes is always computed treating all particles as outgoing.)
With the 2-particle splitting amplitudes defined in eq.~\eqref{eq:split_2}, we can now define 4 ratios of these splitting functions:
\begin{equation}
    \Delta_{\rm I}^{\pm\pm} \coloneq \frac{\splitti^{(4)}_{\pm,\pm}(-p_{1},p_{2},p_{3},-p_{4};\xi_{1},\xi_{2})}{\splitti^{(2)}_{\pm}(-p_{1},p_{2};\xi_{1})\splitti^{(2)}_{\pm}(p_{3},-p_{4};\xi_{2})}\:,
\end{equation}
By specifying $\{h_{5},h_{6}\}$ as $\{+,+\}$ and $\{-,-\}$, we can easily isolate the contributions to $\Delta_{\rm I}^{++}$ and $\Delta_{\rm I}^{--}$, respectively:
\begin{align}
    \Delta_{\rm I}^{++} &=e^{i\delta_{6}}\mathsf{DCL}[R_{6}^{\text{MHV}}(1^{+}4^{+}\to 2^{+}3^{+}5^{+}6^{+})] \:, \label{eq:deltapp}\\
     \Delta_{\rm I}^{--} &=e^{i\delta_{6}}\mathsf{DCL}[R_{6}^{\overline{\text{MHV}}}(1^{+}4^{+}\to 2^{+}3^{+}5^{-}6^{-})]\label{eq:deltamm} \:.
\end{align} 
where $\mathsf{DCL}[R]$ denotes the double collinear limit of the function $R$, $e^{i \delta_{6}}$ arises from the BDS amplitude,  
and $R^{\overline{\text{MHV}}}_6\coloneq A^{\overline{\text{MHV}}}_6/(A_6^{\text{BDS}}A_{\overline{\text{MHV}}}^{\text{tree}})=R_6^{\text{MHV}}$. Then using the argument in Sec.~\ref{sec: general splitting amplitude}, we find 
\[
\Delta_{\rm I}^{++}=\Delta_{\rm I}^{--}=e^{i\delta_{6}}\mathcal{M}(R_{6}^{\text{MHV}})\:. 
\]

To obtain $\Delta_{\rm I}^{+-}$ and $\Delta_{\rm I}^{-+}$, we need the six-point NMHV BDS-subtracted amplitude,  which is conventionally written as 
\begin{align}
    R_{6}^{\text{NMHV}}= ((1)+(4))V_{1}+((1)-(4))\tilde{V}_{1} + \text{cyclic} \:.
\end{align}
Here $V_{i}$ and $\tilde{V}_{i}$ are some parity even and odd transcendental functions, respectively, and $(a)\coloneq[1\ldots\hat{a}\ldots6]$ is the usual $R$-invariant~\cite{Mason:2009qx}. In the double collinear limit, these $R$-invariants behave as
\begin{align}
    (6)\to \frac{-\bar{z}}{1-\bar{z}}(1)\:,\qquad  (2)\to\frac{1}{1-\bar{z}}(1) \:, \\
     (3)\to \frac{-z}{1-z}(4)\:,\qquad  (5)\to\frac{1}{1-z}(3) \:.
\end{align}

Therefore, the $R_{6}^{\text{NMHV}}$ in the double collinear limit can be organized as 
\begin{align}
    \mathsf{DCL}[R_{6}^{\text{NMHV}}]&= (4)\biggl[\frac{1}{1-z}(V_{1}-\tilde{V}_{1}+V_{2}-\tilde{V}_{2}) \nonumber \\ 
    &\qquad\quad +\frac{-z}{1-z}(V_{1}-\tilde{V}_{1}+V_{3}+\tilde{V}_{3})\biggr] \nonumber \\
    &\quad + (1)\biggl[\frac{1}{1-\bar{z}}(V_{1}+\tilde{V}_{1}+V_{2}+\tilde{V}_{2})  \nonumber \\
    &\qquad \quad +\frac{-\bar{z}}{1-\bar{z}}(V_{1}+\tilde{V}_{1}+V_{3}-\tilde{V}_{3})\biggr] 
\end{align}
The effect of the multi-Regge limit on $V_{i}$ and $\tilde{V}_{i}$ is the same as that of the double collinear limit; however, the multi-Regge limit for the process $1^{+}4^{+}\to 2^{+}3^{+}5^{-}6^{+}$ additionally requires $(4)\to 0$, while the multi-Regge limit for the process $1^{+}4^{+}\to 2^{+}3^{+}5^{+}6^{-}$ additionally requires $(1)\to 0$. Meanwhile, the $R$-invariants $(1)$ and $(4)$, together with the Parke-Taylor pre-factor, yield the product of the corresponding tree-level splitting functions and tree amplitudes in this double collinear limit. In this way, we conclude that 
\begin{align}
        \Delta_{\rm I}^{-+} &=e^{i\delta_{6}}\mathcal{M}(R_{6}^{\text{NMHV}}(1^{+}4^{+}\to 2^{+}3^{+}5^{-}6^{+})) \:, \label{eq:deltamp}\\
     \Delta_{\rm I}^{+-} &=e^{i\delta_{6}}\mathcal{M}(R_{6}^{\text{NMHV}}(1^{+}4^{+}\to 2^{+}3^{+}5^{+}6^{-})) \:. \label{eq:deltapm}
\end{align}
Note that, the MHV amplitudes in this double collinear limit only receive contributions from $\Delta_{\rm I}^{++}$, 
while the NMHV amplitudes only receive contributions from $\Delta_{\rm I}^{++},\Delta_{\rm I}^{+-},\Delta_{\rm I}^{-+}$, and they first appear together in the seven-point NMHV amplitudes, all four terms become present starting from the eight-point N$^2$MHV amplitudes. 

The analysis for Region II is similar but now corresponds to the scattering process $1^{+}3^{-}5^{-h_5}\to 2^{+}4^{-}6^{h_{6}}$.

\bibliographystyle{bibliostyle}
\bibliography{main}

\end{document}